\documentclass[prb,floatfix,showpacs,twocolumn]{revtex4}

\usepackage{epsfig}
\usepackage{amsmath}
\usepackage{amssymb}
\usepackage{color}

\newcommand{\nubar}{\nu^{(2D)}}
\newcommand{\Ebar}{E^{(2D)}}
\newcommand{\Ebareff}{\Ebar_{\mathit{eff}}}
\newcommand{\nubareff}{\nubar_{\mathit{eff}}}
\newcommand{\nueff}{\nu_{\mathit{eff}}}
\newcommand{\Eeff}{E_{\mathit{eff}}}

\newcommand{\strain}{\epsilon}
\newcommand{\stress}{\sigma}

\begin{document}

\title{Effective Elastic Moduli in Solids with High Crack Density}
\author{Robert Spatschek}
\affiliation{Interdisciplinary Centre for Advanced Materials Simulation, Ruhr-Universit\"at Bochum, 44780 Bochum, Germany}
\author{Clemens Gugenberger}
\author{Efim Brener}
\affiliation{Institut f\"ur Festk\"orperforschung, Forschungszentrum J\"ulich, 52425 J\"ulich, Germany}

\date{\today}

\begin{abstract}
We investigate the weakening of elastic materials through randomly distributed circles and cracks numerically and compare the results to predictions from homogenization theories.
We find a good agreement for the case of randomly oriented cracks of equal length in an isotropic plane-strain medium for lower crack densities;
for higher densities the material is weaker than predicted due to precursors of percolation.
For a parallel alignment of cracks, where percolation does not occur, we analytically predict a power law decay of the effective elastic constants for high crack densities, and confirm this result numerically.
\end{abstract}

\pacs{46.65.+g, 62.20.mt, 46.25.-y}

\maketitle

\section{Introduction}

The appearance of cracks is an effective mechanism for a mechanical system under load to release elastic energy and to relax towards equilibrium.
It is therefore not surprising that aging processes in a broad class of materials can lead to the emergence of microcracks that weaken a specimen.
They do not necessarily lead to complete failure, but their presence alters the elastic properties of the system. 
For many practical applications it is therefore highly desirable to develop simple, but still precise predictions for the resulting elastic properties of a medium that contains defects, cracks or other inhomogeneities.
Cracked material is just one case in the widely dealt-with topic of physical properties of heterogeneous media \cite{NematNasser,KachanovSegostianov05,Mura91,Bruggeman35}.
The different physical properties to be described encompass conductive, transport and also elastic quantities \cite{FengThorpeGarboczi85,Eshelby57,HashinShtrikman62}.
Often, one starts from a coarse-grained picture and aims to find an effective description for the heterogeneous mixture.
Much effort has been put into the calculation of effective elastic properties of composed media where the constituents have different elastic coefficents \cite{HashinShtrikman63, Giordano03, GiordanoColombo07a}. 

It turns out that the elastic properties of the system depend strongly on the positional and orientational distribution of the inclusions.
Even different loading paths can lead to a different elastic response of the material under investigation \cite{HoriiNemat-Nasser83}.
Therefore, special attention has to be paid to the underlying assumptions of the cavity distribution.
It has been shown that the Hashin-Shtrikman-typ bounds obtained for the bulk modulus of two-phase materials set important restrictions in terms of phase moduli and volume fractions \cite{HashinShtrikman62,HashinShtrikman63}, and improvement to these bounds have to involve considerations of statistical details of phase distributions. 

The starting point for various {\em effective medium theories} is the effect of a single impurity or crack inside an otherwise homogeneous medium.
Also in the following work we will implicitly employ the results of Eshelby \cite{Eshelby57} concerning the elastic fields around and inside a single ellipsoidal inhomogeneity in an infinitely extended, linearly elastic homogeneous solid.
In fact, this ``dilute'' limit for a single imperfection already gives an expression for the effective elastic constants for vanishingly low defect concentration.
Higher concentrations of inhomogeneities can be treated in the framework of self-consistent or differential effective medium theories.
Generally speaking, these approaches make use of the idea, that a medium which already contains inclusions of another ``phase'' can be approximated as a homogeneous material with different material properties, to which then, step by step, additional inhomogeneities are added to reach a finite concentration of them.
The underlying assumption, that all effective properties depend only on the material constants of the pure phases and their volume fractions, is of course only an approximation, and the quality of the theoretical predictions can hardly be controlled.
A careful comparison to either experiments or numerical calculations of the effective properties is therefore highly recommended to judge the quality of the different homogenization methods.

The purpose of the present paper is manifold:
First, it is intended as a numerical check for the analytical estimates for the effective elastic constants.
Obviously, all schemes mentioned above are approximative in nature, and it is one goal of this paper to shed light on the range of applicability of the theoretical models.
We use both finite difference and finite element methods for the numerical investigations, and the comparison to earlier results serves as benchmark for these approaches.
This will be done for the important case of spherical inclusions, since rigorous theoretical statements can be used to test the numerical methods.
This methodological confirmation is essential for the following tests of homogenization theories concerning the weakening of materials through cracks, which is the second main subject of this work.
As will be pointed out, the effect of percolation plays an important role here, and therefore deviations from differential homogenization theories, which predict an exponential weakening of the material, are noticeable already for moderate crack densities.
In this context, the only situation where percolation does not play a role is that of parallel cracks.
The third important subject of this paper is the prediction of effective elastic constants in such a geometry, which surprisingly turn out to decay here according to a power law behavior, in contrast to exponential decays that could be expected from related situations \cite{GiordanoColombo07a}.
It must be pointed out that this fully analytical prediction becomes accurate in the limit of {\em high} crack densities, and is therefore complementary to conventional theories.
The predictions are confirmed by the same numerical methods that have been justified before.

\section{Model verification: Random distribution of spherical holes}

The first system under investigation is that of a two-dimensional isotropic solid in a plane strain situation that contains randomly placed circular holes which are allowed to overlap.
This system has already been investigated numerically by Day et al.~\cite{DaySnyderGarbocziThorpe92}.
We use this scenario to demonstrate the applicability of our numerical method to determine the effective elastic constants.

For $N$ spherical holes of radius $r$ in the solid phase with area $A$, the true void concentration $c$ is related to the void area ratio $\tilde{c}=N\pi r^2/A$ according to the relation
\begin{equation} \label{eq_SphericalHoles5}
c = 1-\exp(-\tilde{c}),
\end{equation}
which takes into account that the circles can overlap. 

Starting from the exact expression for a single inclusion \cite{Eshelby57}, low-density expressions for the effective elastic constants can be derived in terms of the two-dimensional elastic moduli:
\begin{align}
\Ebar_{low} =& \Ebar - 3\Ebar c + {\cal O}(c^2), \label{sphericalholes::eq1} \\
\nubar_{low} =& \nubar + (1 - 3 \nubar)c + {\cal O}(c^2) \label{sphericalholes::eq2}
\end{align}
with the elastic constants $\Ebar, \nubar$ of the solid phase;
this result is attributed to numerous authors \cite{HoriiNemat-Nasser83, SnyderGarbocziDay92}.
We use here the explicit annotation $2D$ to emphasize that the elastic constants are those of a two-dimensional plane strain material, since some peculiarities in the behavior of the effective constants are purely attributed to the dimensionality of representation, as will be elucidated below.
The expressions for conversion between 2D and 3D are given in Appendix \ref{conversion}.

The truncation of the above series after the linear term already provides a low density prediction for the effective elastic constants.
Within this effective medium theory, the elastic modulus $E$ vanishes for $c=1/3$;
however, the true percolation point is\cite{BalbergAndersonAlexanderWagner84,Balberg85} $c_p\approx 0.68$, and only then $E$ should become strictly zero.
This deviation already shows that the effective medium theory loses its predictive power for higher concentrations, underestimating the true stiffness of the material.

Using the above low-density expressions (\ref{sphericalholes::eq1}) and (\ref{sphericalholes::eq2}), we can also derive another approximative model for the elastic constants in the framework of the differential medium theory (see also Appendix \ref{diffhom}).
According to equation (\ref{diffhom::eq3}), we start from
\begin{align}
\frac{d\Ebareff}{dc} &= \frac{-3\Ebareff}{1-c}\\
\frac{d\nubareff}{dc} &= \frac{1 - 3 \nubareff}{1-c}
\end{align}
and obtain as solution
\begin{align}
\Ebareff(c) =& \Ebar(1-c)^3 \label{sphericalholes::eq3},\\
\nubareff  =& \frac{1}{3}-\left(\frac{1}{3}-\nubar \right)(1-c)^3. \label{sphericalholes::eq4}
\end{align}
Apparently, this model predicts ``percolation'' for $c=1$, i.e. if the solid phase disappears completely.
It is obvious that this model therefore must be invalid for high cavity concentrations as well, overestimating the elastic constants of the heterogeneous system.

We note that in both approximative theories the effective elastic modulus does not depend on the Poisson ratio, a behavior that is known to hold exactly \cite{Cherkaev}.

We use a straightforward finite difference method to solve the problem numerically.
In a discretized rectangular system in the $yz$ plane an ``order parameter'' $\phi$ is set to zero at the grid points which are covered by the circles of equal diameter, and $\phi=1$ in the remaining solid.
Then the local elastic modulus is set to $\Ebar(\phi)=\phi \Ebar$, and the elastic equilibrium conditions $\partial \sigma_{ij}/\partial x_j=0$ are solved by relaxation.
The system is strained and the average stress calculated, from which the effective elastic constants can be deduced as follows:
For a system that is strained in $z$ direction and has periodic boundary conditions in $y$ direction, the average strain $\langle \epsilon_{yy}\rangle$ vanishes.
The average diagonal stress components in the system $\langle \sigma_{yy} \rangle$ and $\langle \sigma_{zz} \rangle$ are measured for this plane strain scenario.
Then the effective elastic constants are determined through
\begin{eqnarray}
E_{\mathit{eff}}^{(3D)} &=& \frac{(2\langle \sigma_{yy} \rangle + \langle \sigma_{zz} \rangle)(\langle \sigma_{zz} \rangle-\langle \sigma_{yy} \rangle)}{(\langle \sigma_{yy} \rangle + \langle \sigma_{zz} \rangle) \langle \epsilon_{zz}\rangle}, \\
\nu_{\mathit{eff}}^{(3D)} &=& \frac{\langle \sigma_{yy} \rangle}{\langle \sigma_{yy} \rangle + \langle \sigma_{zz} \rangle},
\end{eqnarray}
where the average strain $\langle \epsilon_{zz}\rangle$ is fixed through the boundary conditions.
We typically used systems sizes of $2048\times 1024$ grid points, with up to 1000 circles with a radius of 20 grid points.
Further details on the elastic solver are presented in \cite{PhaseFieldCrack,CrackGugenberger}.

\begin{figure}
\begin{center}
\includegraphics[width=0.95\linewidth]{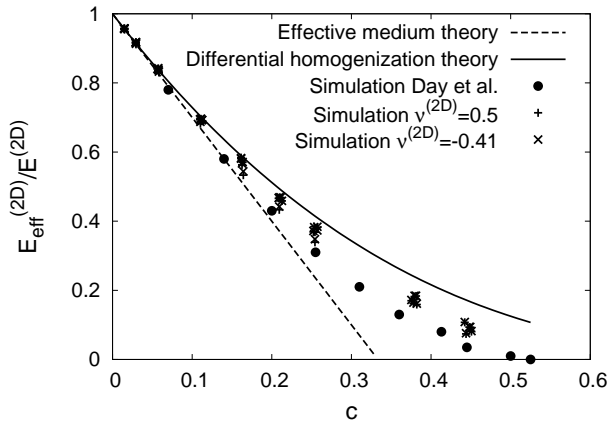}
\caption{Effective elastic modulus as function of the void concentration $c$. The plot shows numerical data for different Poisson ratios, as obtained with the present method, in comparison to numerical results obtained by Day et al. Here we used different distributions of cracks, and evaluated the effective elastic modulus for the two different Poisson ratios using exactly the same arrangement of cracks; the independence of the Poisson ratio is clearly visible.
}
\label{sphericalholes::fig2}
\end{center}
\end{figure}

The dependence of the effective elastic modulus on the concentration as predicted by the theories, see Eq.~\eqref{sphericalholes::eq1} and Eq.~\eqref{sphericalholes::eq3}, and as obtained by numerical simulations is shown in Fig.~\ref{sphericalholes::fig2}.
The independence of $\Eeff$ on the Poisson ratio is clearly visible also in the numerics, where we checked this explicitly for $\nu^{(2D)}=1/2$ and $\nu^{(2D)}=-0.41$ (corresponding to $\nu^{(3D)}=1/3$ and $\nu^{(3D)}=-0.7$ respectively);
the latter case of an auxetic material is sometimes observed e.g.~in foams \cite{Lakes87}, and is here only used as an extreme case to confirm the independence on the Poisson ratio.
In fact, we find that for the same random arrangement of circular holes the elastic constants match.
Since we wanted to obtain a reasonable statistical averaging,  we also performed repeated runs with different initializations.
As we increase the void concentration $c$, one can clearly see that the scattering of the data points increases for higher concentrations, since larger clusters can form which can become comparable to the (finite) system size used in the simulations.
Also, the relaxation time increases strongly with $c$, thus results for higher concentrations are not shown here.
In Ref.~\onlinecite{DaySnyderGarbocziThorpe92}, Day et al. performed simulations based on an elastic spring network formulation for system setups analogous to ours.
The comparison of our numerical results to the simulation data of Day et al. are also included in Fig.~\ref{sphericalholes::fig2}.
The results for the independent numerical approaches are in reasonable agreement. In particular, all sets correctly reproduce the exactly known low density limit $c\to 0$.
For higher concentrations, we obtain a higher effective elastic modulus than Day et al., and we believe that this is a consequence of the considerably larger systems that we used.
\begin{figure}
\begin{center}
\includegraphics[width=0.95\linewidth]{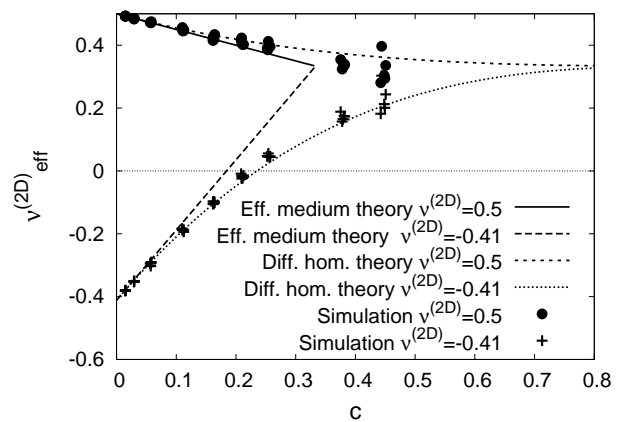}
\caption{Poisson ratio as function of the void concentration $c$. For $\nu = 1/3$, both, the effective medium theory and the differential theory show fairly good agreement with the numerical results. The predictions from the effective medium theory are shown only up to the percolation point $c=1/3$. For negative Poisson ratios, the differential theory coincides much better with the simulations.
}
\label{fig_Spherenueff}
\end{center}
\end{figure}

Similarly, the effective Poisson ratio agrees well with the differential theory, as can be seen in Fig.~\ref{fig_Spherenueff}, especially in the case of a negative Poisson ratio.
Even at the highest densities that were simulated here, we do not observe a noticeable deviation from this homogenization model.

Finally, we briefly remark that the results depend on the dimension of representation.
Conversion of the results for the differential homogenization theory gives according to Eqs.~(\ref{conversion::eq2}) 
\begin{eqnarray}
\Eeff^{(3D)} &=& \Big[ 3E^{(3D)}\;(c - 1)^{3} ( c ( 8\nu^{(3D)} - 2)(c^2-3c+3) - \nonumber \\
&& 3(1 + \nu^{(3D)}) ) \Big] / \Big[(c\;(4\nu^{(3D)}-1)(c^2-3c + 3) - 3)^{2} \nonumber \\
&&\times (\nu^{(3D)}+1)\Big], \\
\nueff^{(3D)} &=& \frac{c\;(4\nu^{(3D)} - 1) (c^2-3c+3) - 3\nu^{(3D)}}{c\;(4\nu^{(3D)}-1)(c^2-3c+3) - 3}.
\end{eqnarray}

In particular, the effective three-dimensional elastic modulus does not have the property of being independent of the Poisson ratio. 
Furthermore, for negative Poisson ratios the effective elastic modulus can first increase if the material is ``weakened'' by spherical holes.
A similar behavior was reported for cracks in Ref.~\onlinecite{GiordanoColombo07a}, and here we see that this effect is rather generic and results mainly from the definition of the elastic constants.
Indeed, this counterintuitive behavior is obviously an artifact of the three-dimensional representation that is already contained in the low density expressions and not related to a specific homogenization scheme.
Already for low concentrations we get
\begin{equation}
\Eeff^{(3D)} = E^{(3D)} \left( 1- \frac{(1-\nu^{(3D)})(8\nu^{(3D)}+3)}{1+\nu^{(3D)}}c \right) + {\cal O}(c^2),
\end{equation}
which can start with a positive slope for negative Poisson ratios.

\section{Random distribution of cracks}

In this section we investigate a random arrangement of cracks in a solid and compare the prediction for the effective elastic constants to numerical simulations.
To that end, we use the same geometry as in Ref.~\onlinecite{GiordanoColombo07a}, where the normal vectors of the planar cracks are located in the $yz$ plane, and they are infinitely extended in $x$ direction.
Therefore, the system becomes again effectively two-dimensional, and we restrict our investigations to a plane-strain scenario.
In the $yz$ plane, all cracks have the same length $L$;
here we assume that the orientation is random and all angles $\theta$ appear with the same probability;
in the notation of Ref.~\onlinecite{GiordanoColombo07a} this means for the orientational order parameter $P=\langle \sin^2\theta\rangle =  1/2$.

We introduce a crack density parameter
\begin{equation}
\alpha = \frac{\pi (L/2)^2 N}{A},
\end{equation}
where $N$ is the number of cracks per area $A$ in the $yz$ plane.
The prediction for the effective (three-dimensional) elastic constants in the framework of the differential homogenization method is for plane strain according to \cite{GiordanoColombo07a}
\begin{align}
\label{eq_EeffGiordano}
\Eeff^{(3D)}  &= \frac{E^{(3D)} [2\nu^{(3D)} + (1-\nu^{(3D)})e^{\alpha}]}{[\nu^{(3D)} + (1-\nu^{(3D)})e^{\alpha}]^2 (1+\nu^{(3D)})}\\
\nueff^{(3D)} &= \frac{\nu^{(3D)}}{\nu^{(3D)} + (1-\nu^{(3D)})e^{\alpha}}\>,
\label{eq_NueffGiordano}
\end{align}
which predicts an exponential weakening of the material with the density parameter $\alpha$.
In particular, the effective medium is still isotropic, since there is no preferred orientation for the cracks, and therefore the elastic properties are still fully described by two elastic constants.

Interestingly, the two-dimensional representation of the above result gives simply
\begin{equation} \label{randomcrack::eq3}
\Ebareff = \Ebar \exp(-\alpha),\qquad \nubareff = \nubar \exp(-\alpha),
\end{equation}
so both constants decay according to a simple exponential decay to zero.
Notice in particular that the effective Poisson two-dimensional ratio also tends to zero, in contrast to the spherical case discussed before, where it approaches $1/3$.
We also mention that here the effective elastic modulus does not depend on the bare Poisson ratio $\nubar$.
Notice that the above conversion implies also that the effect of an increase of stiffness with the crack density for negative Poisson ratios, that was discussed in Ref.~\onlinecite{GiordanoColombo07a}, is indeed an artifact of the three-dimensional representation, similar to the spherical example discussed above.

We note that the limit $\Eeff=0$ is only reached for $\alpha\to\infty$, which means that this theory does not predict percolation.
However, in reality percolation occurs for \cite{PikeSeager74} $\alpha\approx 4.49$, and then a network of cracks penetrates the whole system, thus the true effective modulus vanishes. 
Therefore the differential homogenization method overestimates the true elastic modulus for higher crack densities.

\begin{figure}
\begin{center}
\includegraphics[width=0.95\linewidth]{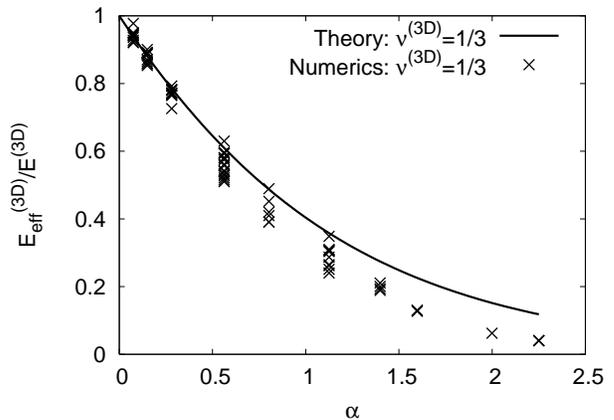}
\caption{Effective elastic modulus as function of the crack density for plane strain loading, $\nu^{(3D)}=1/3$, for several random distributions ($P=1/2$) of cracks of equal length.}
\label{fig_RandomOrientationEeffNu0p3}
\end{center}
\end{figure}

\begin{figure}
\begin{center}
\includegraphics[width=0.95\linewidth]{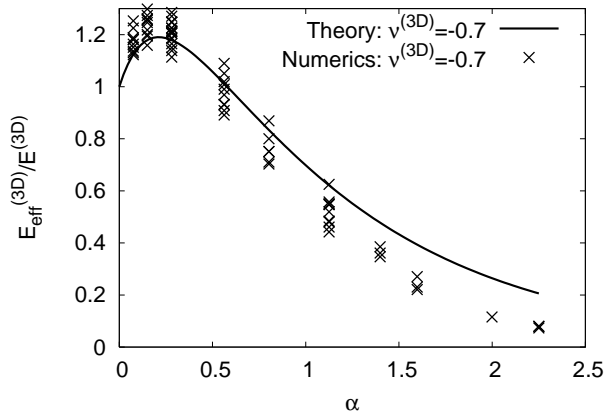}
\caption{Effective elastic modulus as function of the crack density for plane strain loading, $\nu^{(3D)}=-0.7$, for several random distributions ($P=1/2$) of cracks of equal length. The initial stiffness increase predicted by the differential homogenization theory is clearly visible. For higher crack densities, the theory overestimates the effective elastic modulus significantly.}
\label{fig_RandomOrientationEeffNuM0p7}
\end{center}
\end{figure}

To check the quality of the above analytical predictions, we investigated the case of randomly oriented cracks also numerically for plane strain using finite difference relaxation methods, see Figs.~\ref{fig_RandomOrientationEeffNu0p3} and \ref{fig_RandomOrientationEeffNuM0p7}.
For low crack densities, the numerical results agree with the prediction Eq.~\eqref{eq_EeffGiordano}, but for higher values they are indeed systematically lower due to prospective percolation.
\begin{figure}
\begin{center}
\includegraphics[width=0.95\linewidth]{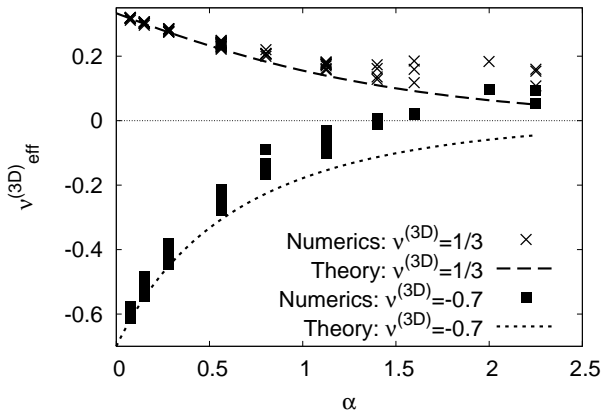}
\caption{Effective Poisson ratio as function of the crack density for plane strain loading for several random distributions ($P=1/2$) of cracks of equal length.}
\label{randomcrack::fig4}
\end{center}
\end{figure}
Nevetheless, the analytical prediction from Ref.~\onlinecite{GiordanoColombo07a} can be considered as a very good approximation at least for crack densities $\alpha < 1$.

We also see good agreement for the Poisson ratio in this range of $\alpha$, see Fig.~\ref{randomcrack::fig4}.
For a negative bare Poisson ratio, here $\nu^{(3D)}=-0.7$, the numerical results seem to indicate that it approaches even a positive value instead of just decaying to zero.

\section{Asymptotic behavior of parallel cracks}
\label{parallel}

From a more general point of view, all setups with random crack orientations have a finite percolation threshold, even if the probability distribution for the choice of the angle is not uniform.
The only exception is the case that all cracks are parallel;
then percolation does not occur.
Thus only here a nontrivial asymptotic behavior exists for high crack densities.
It turns out that for this special case analytical predictions for the effective elastic constants can be made, which become accurate in the limit $\alpha\to\infty$, and in this respect they differ fundamentally from conventional homogenization theories.

\begin{figure}
\begin{center}
\includegraphics[width=1.0\linewidth]{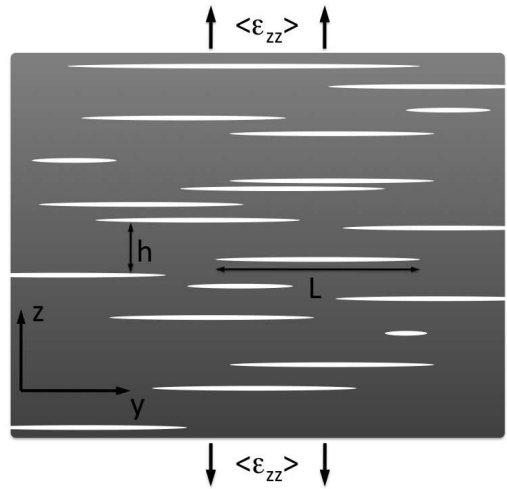}
\caption{Random arrangement of parallel cracks. The average crack length is $L$, the average vertical distance between neighboring cracks $h$.}
\label{parallel::fig1}
\end{center}
\end{figure}

First, it should be noted that in this case the material that is pierced by cracks becomes anisotropic, and we therefore characterize its elastic properties by the tensor $c_{ijkl}^{\mathit{eff}}$, with $\sigma_{ij}= c_{ijkl}\epsilon_{kl}$.
If we assume that in the $yz$ plane all cracks are aligned in $y$ direction (see Fig.~\ref{parallel::fig1}), it is immediately clear that e.g. $c_{yyyy}^{\mathit{eff}} = c_{yyyy}$, since a pure stretching in $y$ direction does not open the cracks;
hence the strain tensor is homogeneous in the material and unaffected by the cracks.

For low crack densities, the effective elastic constants were calculated in Ref.~\onlinecite{GiordanoColombo07a}, and in particular we get
\begin{equation} \label{parallel::eq1}
c_{3333}^{\mathit{low}} = (1-\nu^{(3D)})[1+2\alpha P] D^{-1} E^{(3D)}
\end{equation}
with
\begin{eqnarray*}
D &=& [4\alpha^2 P(1-P)(1-\nu^{(3D)})^2 + 2(1-\nu^{(3D)})^2\alpha + \\
&& + 1 - 2\nu^{(3D)} ] (1+\nu^{(3D)})
\end{eqnarray*}
and $P=0$ for the parallel arrangement.

\begin{figure}[htpb]
	\begin{center}
		\includegraphics[width=1.0\linewidth]{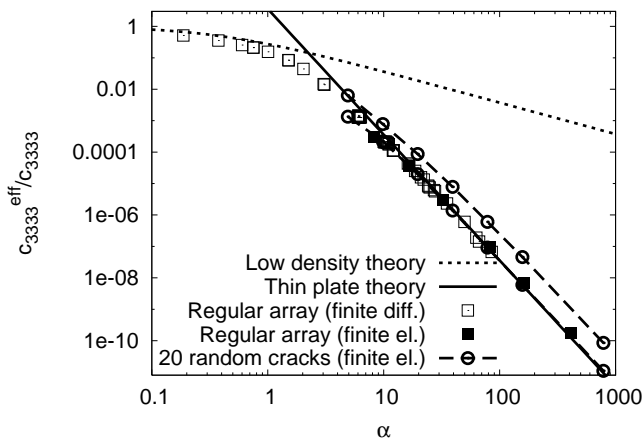}
		\caption{Scaling behavior of the effective elastic constant $c_{3333}^{\mathit{eff}}$ as a function of the crack density $\alpha$ for a parallel arrangement of cracks in logarithmic representation. For a regular arrangement of cracks, the agreement of the numerical simulations with thin plate theory is excellent. If the cracks are placed at random positions, they still exhibit the same power law scaling behavior.}
		\label{fig_ScalingParallelCracks}
	\end{center}
\end{figure}

We start with looking at high crack densities, $\alpha \to \infty$: two different lengthscales are important for a complete description of the problem at hand, the length $L$ of the cracks and the average vertical distance $h$ between them. For high crack densities $\alpha$, the vertical distance $h$ between neighboring cracks is smaller than the average crack length $L$, and the relation between the two characteristic length scales can be given through $\alpha$ only, so we obtain $h \sim L/\alpha$. If the cracked body is subjected to tensile loading perpendicular to the cracks, the solid regions between two cracks can be understood as a thin bent plate of a width proportional to $L$ and thickness $h$. The opening of the cracks is the displacement $u_z$. The stress of a thin bent plate scales as \cite{LandauLifshitz, BrenerMuellerKrumbhaarSpatschek01}
\begin{equation}
	\stress_{zz} \sim \frac{Eh^3}{1-\nu^2} \frac{\partial^4 u_z}{\partial y^4} \>.
\end{equation}
With this equation, it follows readily that the average stress and the opening $u_z$ have to scale like
\begin{equation}
	\langle u_z \rangle \sim \langle \stress_{zz} \rangle \frac{(1-\nu^2) L^4}{E h^3}\>.
	\label{Eq_UsScaling}
\end{equation}
The total displacement is distributed among the opening of all cracks, which relax the material around them. Since for this loading all other average strain components are small \cite{LandauLifshitz}, the average strain $\langle \strain_{zz} \rangle$ is simply given by 
\begin{equation}
	\langle \strain_{zz} \rangle = \frac{\langle u_z \rangle}{h}.
\end{equation}
Plugging this into Eq.~\eqref{Eq_UsScaling}, we finally obtain for the case $\alpha \gg 1$
\begin{equation}
	\langle \stress_{zz} \rangle \sim \langle \varepsilon_{zz} \rangle \frac{E}{(1-\nu^2) \alpha^4}.
\end{equation}
In other words, the relevant elastic constant 
\begin{equation}
	c_{3333}^{\mathit{eff}} = \frac{\langle \varepsilon_{zz} \rangle}{\langle \stress_{zz} \rangle} 
\end{equation}
decays by a power law, 
\begin{equation} \label{parallel::eq2}
c_{3333}^{\mathit{eff}} \sim c_{3333} \alpha^{-4}.
\end{equation} 
We note that this scaling behavior holds also for situations where the cracks can have unequal lengths, distributed around the mean value $L$;
details of the distribution function can affect only the numerical prefactor of the above prediction in the limit $\alpha\to\infty$.
In addition, we also performed  simulations for regular arrays of cracks.
Also, we checked numerically that the scaling behavior holds for random parallel arrangements of cracks; the results can be seen in Fig.~\ref{fig_ScalingParallelCracks}.
This graph shows the results for the low density theory, the asymptotic behavior and numerical simulation data from both finite difference and finite element methods \cite{freefem}.
We used different arrangements of cracks to illustrate the scaling behavior:
First, we took a regular arrangement of cracks, where we can rigorously calculate the effective elastic constants for $\alpha\to\infty$; 
this is shown in Appendix \ref{regular}.
Due to the spatial periodicity it is sufficient to consider a system with only a few cracks.
We clearly see that both finite difference and finite element calculations give the same result.
The finite element method is computationally more efficient than the simple relaxation solver;
however, the geometrical description is easier with finite differences, since e.g.~intersections with boundaries (or overlaps of cracks for the random orientation case, as discussed in the preceding section) do not require a separate treatment.
To get clear predictions for the scaling behavior as function of the crack density $\alpha$, we randomly place the cracks in the system and solve the elastic problem by finite element methods.
Then we change the value of $\alpha$ by rescaling the height of the system, which means that the arrangement of cracks is the same for all points on one curve.
The correct scaling behavior is demonstrated here for a relatively small system with only $N=20$ cracks.
Obviously, the specific results depend then on the configuration, and only for $N\to\infty$ these discrepancies between different arrangements would disappear.
However, the results show, that the scaling holds for each configuration (shown here for two cases), and therefore it must be correct also for the true ensemble average in an infinitely large system.

The results, in particular the finite difference data for small $\alpha$ show the crossover between the low density prediction (\ref{parallel::eq1}) and the asymptotic behavior (\ref{parallel::eq2}).
For the latter, the numerical prefactor was chosen such that it matches the particular case of regular cracks ($g=1/2$), as explained in Appendix \ref{regular}.

\section{Summary and Conclusion}

We investigated numerically the effective elastic constants for isotropic plane strain media with spherical holes, randomly oriented and parallel cracks.
In all cases we find a good agreement with predictions from different homogenization theories, with a better performance of differential media theories.
The results show clear deviations from the approximative theories, which are strictly valid only for low inclusion densities, since they do not correctly account for effects which go beyond mean-field approximations.
In particular, all discussed models do not correctly take into account percolation, which should lead to a sharp drop of the effective elastic modulus.
The only case where percolation does not occur is that of parallel cracks.
By scaling arguments we derived analytically the scaling behavior of effective elastic constants in the limit $\alpha\to\infty$ and obtain a power law decay with the crack density.
This new prediction was confirmed numerically using finite-difference and finite-element methods.
We note that this prediction is complementary to conventional homogenization theories, as it becomes accurate for {\em increasing} crack densities.
Even though the effective elastic constants are already low in this regime, the obtained results are therefore of principal interest and raise the question whether explicit solutions for other situations with high inclusion density are also possible.

\begin{acknowledgments}
This work has been supported by the German-Israeli Foundation.
R.~S.~would like to acknowledge the financial support from the industrial sponsors of ICAMS, ThyssenKrupp Steel AG, Salzgitter Mannesmann Forschung GmbH, Robert Bosch GmbH, Bayer Materials Science AG, Bayer Technology Services GmbH, Benteler AG and the state of North-Rhine-Westphalia.
\end{acknowledgments}

\appendix

\section{Conversion between two- and three-dimensional representation}
\label{conversion}

As already mentioned above, the dimensionality can play a role for the effective elastic constants.
We can convert the elastic constants of a two-dimensional setup to an equivalent three-dimensional plane strain situation.
The defining equation is Hooke's law,
\begin{equation} \label{conversion::eq1}
\epsilon_{ij} = \frac{1}{E} \left[ (1+\nu) \sigma_{ij} - \nu \delta_{ij} \sigma_{kk} \right]
\end{equation}
which holds for both 3D and 2D;
the difference is that in the first case all indices run over $x, y, z$, in the second only over $y, z$.
In a plane strain 3D configuration, $\epsilon_{xx}=0$, we have $\sigma_{xx}=\nu(\sigma_{yy}+\sigma_{zz})$, whereas this stress component does not appear in 2D.
Hence the conversion rules for the elastic constants are given by
\begin{equation} \label{conversion::eq2}
E^{(3D)} = \Ebar \frac{1+2\nubar}{(1+\nubar)^2} \>, \qquad \nu^{(3D)} = \frac{\nubar}{1+\nubar},
\end{equation}
which follows directly from Hooke's law (\ref{conversion::eq1}).

\section{Differential Homogenization Method}
\label{diffhom}

Let a system of dimensionless ``volume'' $V_0 = 1$ contain inclusions of a second phase, characterized by  the initial concentration (volume fraction) $c_0$ , which in turn means that the concentration of the first phase is $1-c_0$.
Now, a volume $dc_0$ of the second phase to the original volume $V_0 = 1$ is added, leading to a total volume of $V = 1 + dc_0$.
The total volume of phase two has increased to $c_0 + dc_0$, resulting in a total volume fraction of
\begin{equation} \label{diffhom::eq1}
c = \frac{c_0 + dc_0}{1 + dc_0} = c_0 + (1-c_0 )dc_0 + {\cal O}(dc^2_0).
\end{equation} 
The change of concentration of the second phase is therefore $dc = (1-c_0 )dc_0$. 
Let $M_{\mathit{eff}}$ denote a complete set of effective elastic constants (or other quantities of interest).
In the framework of the homogenization methods used here, this set should depend on the properties of the pure phases and the concentration, $M_{\mathit{eff}} = F (M_1 , M_2 , c)$ with a universal function $F$ and the obvious relation
\begin{equation} \label{diffhom::eq2}
M_{\mathit{eff}}= F (M_{\mathit{eff}}, M_2, c = 0).
\end{equation} 
In the framework of the differential homogenization method the increase of the amount of the new phase from $c$ to $c+dc$ is interpreted as the addition of the amount $dc_0$ to the already homogenized medium with properties $M_{\mathit{eff}}$.
Hence we obtain
\begin{eqnarray*}
M_{\mathit{eff}} &+& dM_{\mathit{eff}} = F(M_1, M_2, c+dc) = F(M_{\mathit{eff}}, M_2, dc_0) \\
&=& M_{\mathit{eff}} + \left. \frac{\partial F(M_1, M_2, c)}{\partial c} \right|_{c=0, M_1=M_{\mathit{eff}}(c)} \frac{dc}{1-c}.
\end{eqnarray*}
since in the second step the change of concentration is $dc_0/(1+dc_0)=dc_0 + {\cal O}(dc_0^2)$;
in the last step the relation (\ref{diffhom::eq2}) was used.
Here it is important to note that after the differentiation first $c$ has to be set to zero, and only then $M_1=M_{\mathit{eff}}(c)$ to be inserted.
Therefore, we immediately obtain the fundamental equation
\begin{equation} \label{diffhom::eq3}
\frac{dM_{\mathit{eff}}}{dc} = \frac{1}{1-c} \left. \frac{\partial F(M_1, M_2, c)}{\partial c} \right|_{c=0, M_1=M_{\mathit{eff}}(c)}.
\end{equation}
For slit-like cracks, the volume fraction is zero, and therefore the prefactor $(1-c)^{-1}$ disappears and $c$ is replaced by the density parameter $\alpha$.

\section{Regular array of cracks}
\label{regular}

\begin{figure}
\begin{center}
\includegraphics[trim = 1cm 6.5cm 8cm 0.5cm, clip=true, width=0.9\linewidth]{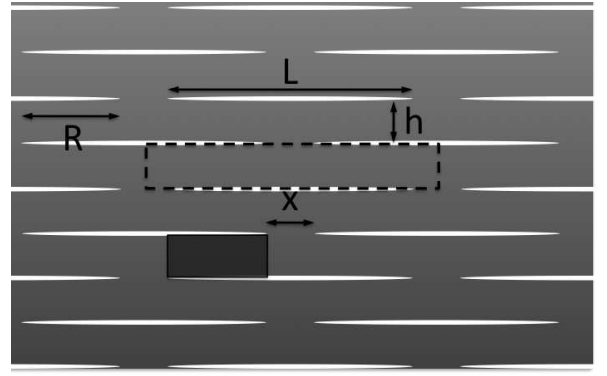}
\includegraphics[trim = 0cm 18cm 19cm 0.0cm, clip=true, width=0.7\linewidth]{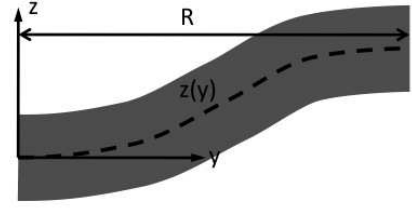}
\caption{Top: Sketch of the regular array of cracks that is used both for analytical calculations and numerics. The dashed rectangle is the ``periodic unit cell'' in which the elastic problem is solved numerically. The dark box visualizes the plate that is bent under the applied load, which is shown in the lower panel. The deformation of the neutral fiber is denoted by $z(y)$.}
\label{regular::fig1}
\end{center}
\end{figure}

To make the preceding scaling arguments in section \ref{parallel} more explicit, we discuss here a regular arrangement of cracks, as depicted in Fig.~\ref{regular::fig1} and solve this problem exactly in the limit $\alpha\to\infty$.
The idea is that the displacement, which is applied to the sample is mainly stored in the opening of the cracks, and the material in between is only slightly stretched.
The region between adjacent cracks behaves then as a bent plate (see dark region in Fig.~\ref{regular::fig1}), which is thin in the limit $R \gg h$.
We note that for this regular arrangement the plate length $R$ appears here as additional parameter, which is related to the gap distance $x$ by $L=2 R +x$;
again, $L$ is the crack length which is now assumed to be exactly the same for all cracks.
Therefore, the additional dimensionless parameter $g=x/R$ remains in the final solution, whereas for an irregular arrangement of cracks it would be determined statistically;
finally, it enters only into the numerical prefactor of the effective elastic constants.

For the given geometry, the area that is occupied by a single crack, $N=1$, is $A=(L+x)h$.
Therefore, the crack density is
\begin{equation} \label{regular::eq1}
\alpha = \frac{\pi}{2} \frac{(1+g/2)^2}{1+g} \frac{R}{h}.
\end{equation}
The bending of the thin plate is described by the equation $z''''(y)=0$, since the upper and lower surfaces are stress free \cite{LandauLifshitz}.
Each plate is displaced by $z(R) = \langle \epsilon_{zz} \rangle h$, since the total displacement is equally distributed among all crack openings.
Together with the symmetry conditions $z'(0)=z'(R)=0$ and the reference value $z(0)=0$, we obtain for the coefficients of the general solution $z(y)=ay^3 +by^2+cy +d$ the values $b=3\langle \epsilon_{zz}\rangle h/R^2$ and $a=-2b/3R$.
The force per unit length in $x$ direction that is required to bend the plate by the given amount is given by \cite{LandauLifshitz}
\begin{equation} \label{regular::eq2}
F = -\frac{E h^3}{12(1-\nu^2)} z''' = \frac{E h^4 \langle \epsilon_{zz}\rangle}{(1-\nu^2)R^3},
\end{equation}
and thus the average stress in vertical direction
\begin{equation} \label{regular::eq3}
\langle \sigma_{zz} \rangle = \frac{F}{x+R} = \frac{E \langle\epsilon_{zz}\rangle}{1-\nu^2} \left( \frac{\pi}{2} \right)^4 \frac{(1+g/2)^8}{(1+g)^5}\alpha^{-4}
\end{equation}
From Hooke's law for the effective medium, $\langle \sigma_{zz}\rangle  = c_{3333}^{\mathit{eff}} \langle \epsilon_{zz}\rangle + \ldots$ follows
\begin{equation} \label{regular::eq4}
c_{3333}^{\mathit{eff}} = \frac{E}{1-\nu^2} \left(\frac{\pi}{2}\right)^4 \frac{(1+g/2)^8}{(1+g)^5} \alpha ^{-4}.
\end{equation}
The bare elastic constant $c_{3333}$ is related to the isotropic moduli by
\begin{equation}
c_{3333} = \frac{E(1-\nu)}{(1+\nu)(1-2\nu)},
\end{equation}
and hence get get asymptotically for $\alpha\to\infty$
\begin{equation}
\frac{c_{3333}^{\mathit{eff}}}{c_{3333}} = \frac{1-2\nu}{(1-\nu)^2} \left(\frac{\pi}{2}\right)^4 \frac{(1+g/2)^8}{(1+g)^5} \alpha ^{-4}.
\end{equation}


\end{document}